\documentclass[aps,prb,preprint,unsortedaddress,showpacs]{revtex4}
\usepackage{amssymb}
\usepackage{graphicx}
\usepackage{color}

\begin{document}

\title{Lattice Properties of PbX (X = S, Se, Te): Experimental Studies and \textit{ab initio}
Calculations Including Spin-Orbit Effects}
\author{A.H. Romero}
\affiliation{CINVESTAV, Departamento de Materiales, Unidad
Quer$\acute{e}$taro, Quer$\acute{e}$taro, 76230, Mexico}

\author{M. Cardona}
\author{R. K. Kremer}
\email[Corresponding author:~E-mail~]{r.kremer@fkf.mpg.de}
\author{R. Lauck}
\author{G. Siegle}
\affiliation{Max-Planck-Institut f{\"u}r Festk{\"o}rperforschung,
Heisenbergstr. 1, D-70569 Stuttgart, Germany}

\author{J. Serrano}
\affiliation{ICREA-Dept. F\'{\i}sica Aplicada, EPSC, Universitat Polit\'{e}cnica de Catalunya,
Av. Canal Ol{\'{\i}}mpic 15, 08860 Castelldefels, Spain}

\author{X. C. Gonze}
\affiliation{Unit\'{e} de Physico-Chimie et de Physique des
Mat\'{e}riaux Universit\'{e} Catholique de Louvain\\
B-1348 Louvain-la-Neuve, Belgium}

\date{\today}

\begin{abstract}
During the past five years the low temperature heat capacity of
simple semiconductors and insulators has received renewed attention.
Of particular interest has been its dependence on isotopic masses and
the effect of spin- orbit coupling in \textit{ab initio} calculations.
Here we concentrate on the lead chalcogenides PbS, PbSe and PbTe.
These materials, with rock salt structure,
have different natural isotopes for both cations and anions, a fact that
allows a systematic experimental and theoretical study of isotopic effects
e.g. on the specific heat. Also, the large spin-orbit splitting of the 6$p$ electrons
of Pb and the 5$p$ of Te allows, using a computer code which includes spin-orbit
interaction, an investigation of the effect of this interaction on the phonon
dispersion relations and the temperature dependence of the specific heat and on
the lattice parameter. It is shown that agreement between measurements and
calculations significantly improves when spin-orbit interaction is included.
\end{abstract}

 \maketitle


\email{M.Cardona@fkf.mpg.de}

\section{Introduction}

Considerable effort has been spent recently in the investigation of the heat
capacity $C$ of semiconductors and insulators in the region below the Debye temperature $\theta_D$ ,
in particular  around  0.1 $\theta_D$ , where strong deviations from Debye's $T^3$  power law take place.
The availability of stable isotopes has also enabled the investigation of the dependence of $C$
on one isotopic mass (in monatomic crystals such as C (Ref. \onlinecite{Cardona2005}),
Si (Ref. \onlinecite{Gibin2005}), Ge (Ref. \onlinecite{Sanati2004,Schnelle2001}),
Sb (Ref. \onlinecite{Serrano2008}), Bi (Ref. \onlinecite{Diaz2007PRL})) or on the isotopic mass of
each constituent in polyatomic compounds, such as GaN (Ref. \onlinecite{Kremer2005}) or Zn0 (Ref. \onlinecite{Serrano2006}),
whereas the development of efficient
computer codes for electronic band structure calculations has made possible \textit{ab initio}
calculations of the phonon dispersion relations and the temperature
dependence of the specific heat.

Although experimental and calculated results for $C_v$ $\approx$   $C_p$  (Ref. \onlinecite{Cardona2007})
agree usually reasonable well, it
has been recently discovered that considerable discrepancies exist when heavy constituent atoms are
present, e.g. Bi (Ref. \onlinecite{Diaz2007PRL}), Pb in PbS (Ref. \onlinecite{Cardona2007}),
if the Hamiltonian used for the calculation of the electronic
structure does not include spin-orbit (s-o) interaction. By performing \textit{ab initio}
calculations in which the s-o interaction is switched on and off, it has been rather
conclusively shown that this interaction "softens" the phonon frequencies (Ref. \onlinecite{Murray2007,Diaz2007})
and thus increases the low-temperature maximum found in $C_v/T^3$ versus $T$.\cite{Diaz2007PRL}

In this article we present \textit{ab initio} calculations of the phonon dispersion
relations of PbS, PbSe and PbTe based on the electronic band structure obtained with the ABINIT code.\cite{Abinit}
This program enables the inclusion of s-o interaction, separately or jointly for the cation or
anion constituents, thus making it possible to separate the s-o contributions to the dispersion relation
and the heat capacity. The ABINIT program determines  the dynamical matrix elements by perturbation
theory. Once these elements are known, it is a simple task to calculate the dispersion relations for
a given set of isotopic masses and thus to determine isotopic effects on the dispersion relations
and the specific heat. Because of the importance of s-o interaction, we have performed the
isotope effect calculations only with s-o interaction included. Within the range of stable
isotopes available in nature (and accessible to our budget) only mass changes of a few percent
are possible: the resulting variations in the physical properties are thus linear in the mass
changes. Results for PbS have already been published in Ref.\onlinecite{Cardona2007}.
At that time, the ABINIT
code did not properly include s-o interaction for diatomic polar compounds. We therefore present
here similar results including s-o interaction: the inclusion of this interaction
reduces the discrepancy between calculated and experimental values
of $C_{v,p}$ (29\%) by a factor of 2. A similar reduction is found for PbSe,
whereas for PbTe the calculated values almost exactly agree with the
measured data. In this latter case the effect of the s-o interaction on
the maximum of $C_v/T^3$ is 18\% (we recall that the s-o interaction softens the phonons and correspondingly
increases $C_v/T^3$ at the maximum). Globally, the inclusion of s-o interaction also improves the agreement
between the calculated and the measured (by inelastic neutron scattering, INS) phonon
dispersion relation, except for the LO phonons in the vicinity of the $\Gamma$-point of the
 Brillouin zone (BZ). At this point, the long range electric fields associated with the
LO phonons lead to some convergence problems which we have not been  able to
avoid completely.

In Ref. \onlinecite{Cardona2007}, and in  Ref. \onlinecite{Etchegoin2008}
we also reported densities of one- and two-phonon (with the same
$k$-value both, as required to compare with optical spectroscopy results) states, calculated for PbS in the
absence of s-o interaction. We present here similar two-phonon spectra obtained for
the three lead chalcogenides including s-o interaction.
The features obtained in the density of one-phonon states, and their projections on the constituent
atoms, are useful (and very instructive) for the interpretation of the isotope effects.

We have also looked at the effect of s-o interaction on the lattice parameter $a_0$ as obtained by energy
minimization. This effect is rather small ($\sim$ 0.2\%). It does not help to reduce the discrepancy between
the calculated and the measured values, which is about 2\%.

\section{Theoretical Details}

The calculations of the dispersion relations and the specific
heat were performed with the dynamical matrix obtained from the LDA electronic structure calculations
using the ABINIT code.\cite{Abinit} Hartwigsen-Goedecker-Hutter relativistic separable dual-space pseudopotentials were used.\cite{Hartwigsen1998} We performed checks with LDA and GGA functionals and convinced ourselves that no significant differences resulted; we therefore used LDA functionals throughout. For the Pb pseudopotential we also checked that our conclusions did not depend on the inclusion of 5$d$ electrons in the valence bands. This allows us to use the pseudopotential as it is given in the ABINIT website. The implementation of the s-o term has been discussed in Ref. \onlinecite{Diaz2007}. We investigated cell parameter convergence as a function of energy cutoff and $k$-grid mesh. With the cutoff at 60 Hartree the total energy is converged up to 0.5 meV and stresses are lower than 0.006 GPa. The BZ is sampled
using a 6 $\times$ 6 $\times$ 6 Monkhorst-Pack grid.\cite{Monkhorst1976} Technical details of the calculations of the phonon dispersion relations can be found in Ref. \onlinecite{Gonze1997a} and \onlinecite{Gonze1997b}.
Prior to obtaining the dynamical matrix the lattice parameter $a_0$ was
optimized through minimization of the total energy. In the calculations we used this parameter and not the
experimental one.

After obtaining the dispersion relations, the phonon free energy $F$ was calculated with the
integral given in Eq. (2) of Ref. \onlinecite{Cardona2007}.
The specific heat was obtained with the expression:

\begin{equation}
C_v =-T\left(\frac{\partial^2 F}{\partial T^2}\right)_v.\label{eq1}
\end{equation}

After calculating the dynamical matrix elements we diagonalized
the Hamiltonian for two different sets of masses (either cation or anion)
differing by about 5\%. The resulting dispersion relations
were then used to calculate the derivatives of $C_v$  with respect to either mass.
The logarithmic derivatives were also calculated because they can be related to
the corresponding derivative with respect to temperature. The appropriate relation
is given in Eq. (4) of Ref. \onlinecite{Cardona2007} for the case of a monatomic crystal. For diatomic crystals
such relation is not as simple since both, derivatives of $C_v$ versus the
masses of cation and anion must be added in order to obtain the temperature derivative:

\begin{equation}
\frac{d \ln (C_{p,v}/T^3)}{d \ln M_{\rm Pb}} + \frac{d \ln (C_{p,v}/T^3)}{d \ln M_{\rm X}}=  \frac{1}{2}\,\,(3 + \frac{d
\ln (C_{p,v}/T^3)}{d \ln T}). \label{eq2}
\end{equation}

\noindent where X = S, Se or Te, respectively.

When calculating numerically the derivatives above at very low temperatures ($\leq$ 5K)
one encounters a convergence problem because of the small values of  $C_v$ and $T$.
Fortunately, the limit $T \rightarrow 0$ can be obtained analytically by using the Debye-$T^3$ approximation. The corresponding expressions are found in Eq. (5) of Ref. \onlinecite{Cardona2007}.

\section{Experimental Procedure}

Samples of  PbX (X = S, Se, Te) were prepared by first reacting the
corresponding pure elements and then subliming the product in an
argon atmosphere.

In order to purify the lead isotope oxide layers on the metal pieces were removed by
etching in diluted nitric acid and, in the case of natural lead, by
cutting. Subsequently, the pieces were melted in silica
ampoules in hydrogen or argon. Remaining oxide
stuck to the ampoule wall after rocking and rolling the droplets at temperatures from 400 to 650$^o$C.
The chalcogens were purified by sublimation and separation of the
ampoule portion containing the residues. In the case of $^{130}$Te,
we first reduced $^{130}$TeO with sulfur.\cite{Gehlen1940}
The synthesis was performed in argon by increasing the temperature to $\sim$650 $^o$C of the chalcogen.
In the sealed silica ampoule crystal growth took place by sublimation at temperatures of 750 to 850 $^o$C under excess of the corresponding chalcogen.
This resulted in a cabbage-like growth of the compound on the lead.
A few platelets up to 4 $\times$ 4 mm$^2$ were obtained during 1 to 2 weeks.\cite{Szczerbkow2005}
The preparation and some properties of the PbS samples are described
in some of the measured PbS samples were natural galena crystals.\cite{Sherwin2005}

The heat capacities were measured between 2 and 280K with a PPMS
system (Quantum Design, 6325 Lusk Boulevard, San Diego, CA.) as
described in detail Ref. \onlinecite{Serrano2006}.

We also measured natural galena crystals.

\section{Dispersion Relations}
\begin{figure}[tbph]
\includegraphics[width=8cm ]{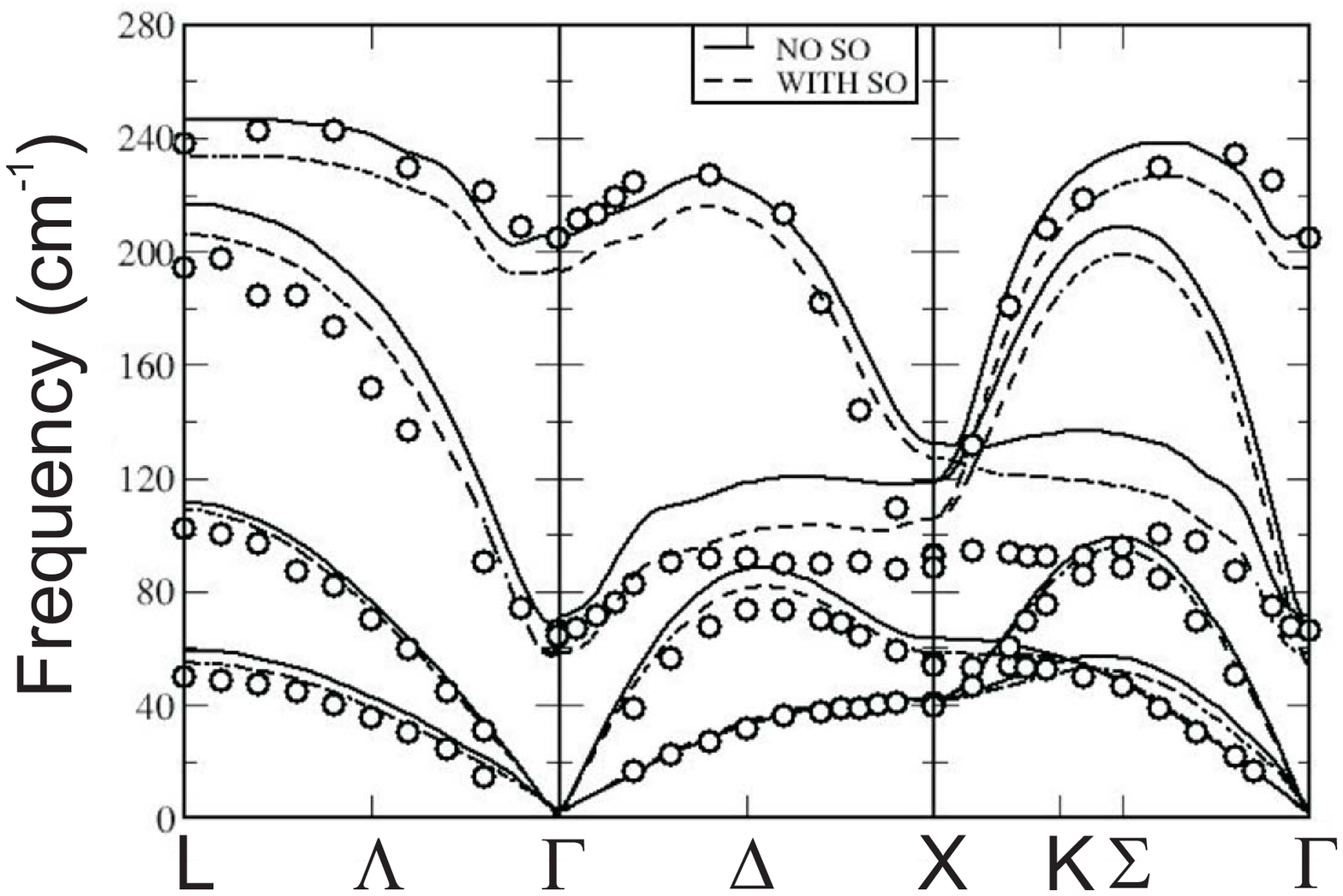}
\caption{Phonon dispersion relations of  PbS with natural isotope composition of Pb and S
calculated with
and without s-o splitting within the harmonic approximation. The points
were obtained by INS at 300 K (From Ref. \onlinecite{Elcombe1967}).
Typically, anharmonic
effects should have lowered them by $\sim$ 2 cm$^{-1}$.} \label{fig1}
\end{figure}

The phonon dispersion relations calculated along three high symmetry directions of
the BZ ( [111], [100], [011]) for PbS are shown in Fig. \ref{fig1} together with INS
data of Elcombe.\cite{Elcombe1967}
The calculations without s-o coupling were already published in Ref. \onlinecite{Cardona2007}.
Note that the s-o
interaction lowers all phonon frequencies, bringing the calculations in better agreement
with the experiments except for the LO band where, on the average, the agreement is
similar. As we shall see later, the decrease in the calculated frequencies raises he
maximum in $C_v/T^3$ , thus improving agreement with experimental data.

\begin{figure}[tbph]
\includegraphics[width=8cm ]{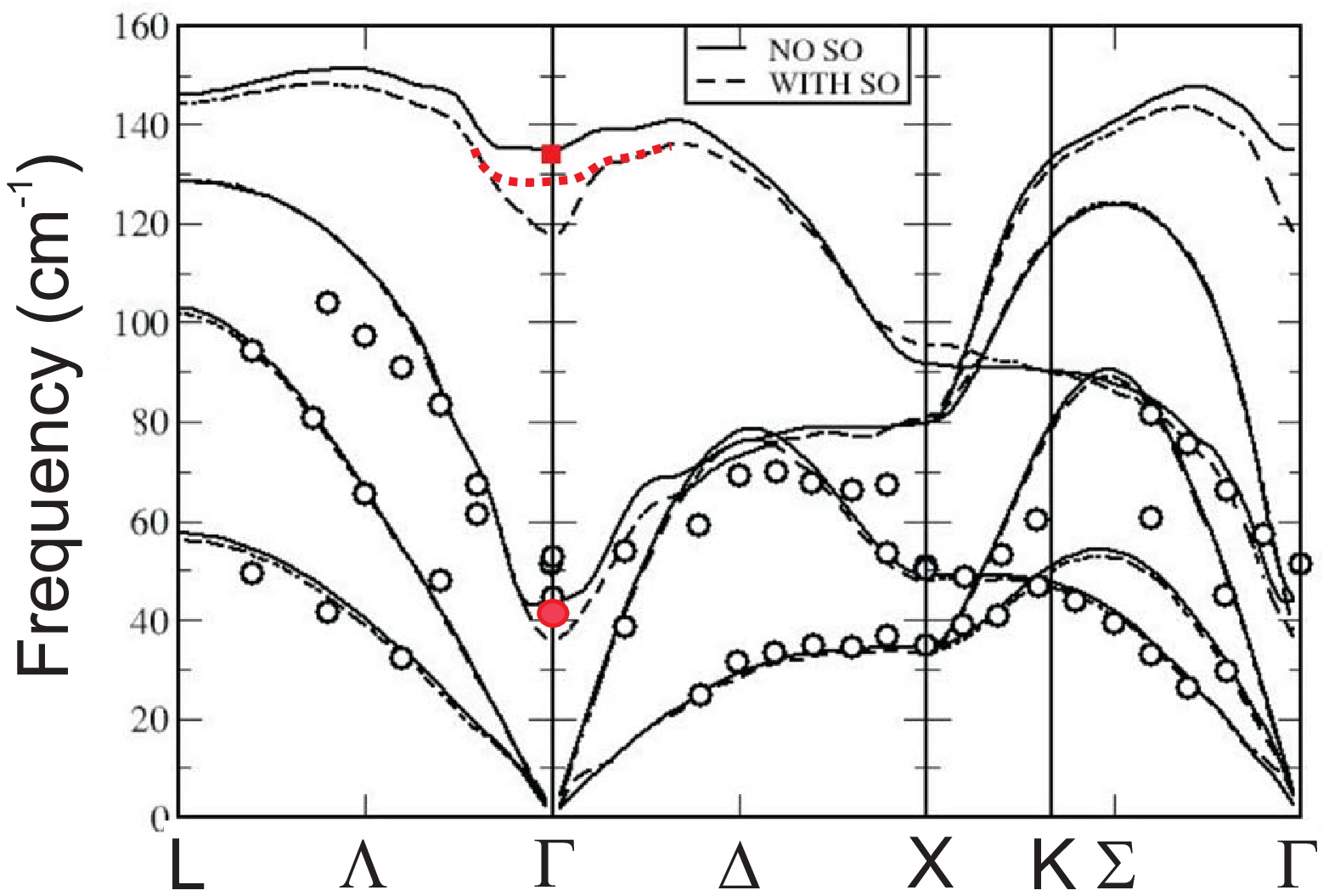}
\caption{(color online) Phonon dispersion relations of  PbSe with natural isotope composition of Pb and Se
calculated with
and without s-o splitting within the harmonic approximation. The circles ($\circ$)
were obtained by INS at 300 K (From Ref. \onlinecite{Vijayraghavan1963}). The (red) square was obtained
by tunnel spectroscopy at 4.2 K.\cite{Hall1961}
Typically, anharmonic
effects should have lowered them by $\sim$ 2 cm$^{-1}$. The (red) closed circle
was obtained by ir transmission at 1.4K.\cite{Burstein1964} The (red) dotted line isuggests how the LO band should
look like (for more details see text).} \label{fig2}
\end{figure}

\begin{figure}[tbph]
\includegraphics[width=8cm ]{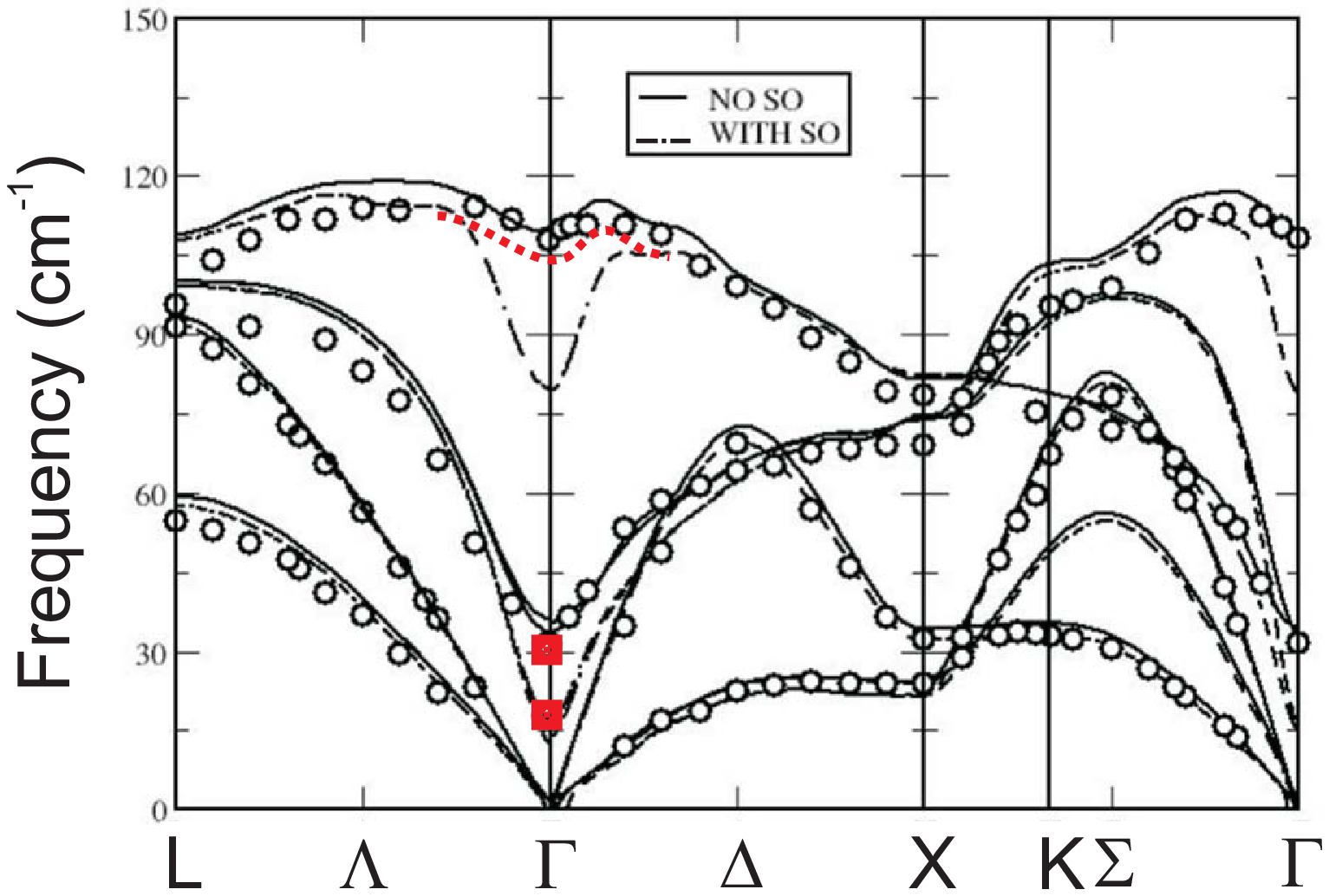}
\caption{(color online) Phonon dispersion relations of  PbTe with natural isotope composition of Pb and Te
calculated with
and without s-o splitting within the harmonic approximation. The (black) circles
were obtained by INS at 297 K (From Ref. \onlinecite{Cochran1966}).
The (red) squares were obtained by optical spectroscopy at 300K (32 cm$^{-1}$) and 5K (18 cm$^{-1}$).\cite{Bauer1978}
The (red) dotted line suggests how the LO band should
look like (for more details see text)} \label{fig3}
\end{figure}

Figure \ref{fig2} and \ref{fig3}
display the dispersion relations of PbSe and PbTe respectively,
calculated with and without s-o coupling, together with experimental data.
For both materials we also observe the decrease of phonon frequencies when the
s-o interaction is taken into account. For PbSe only few experimental (INS) data points are
available.\cite{Vijayraghavan1963}
Figure \ref{fig2} provides some  indication that also in this case the calculations with s-o coupling come closer
to the experimental data than those without.

Figure \ref{fig3} shows calculated (with and without s-o interaction) and measured
(by INS at 297K)  phonon dispersion relations of PbTe. Again we observe a lowering of nearly all frequencies induced
by taking into account s-o interaction, which also brings the calculations, in most cases,
closer to the experimental data. The measured TO frequencies are known to be strongly
renormalized downwards by the anharmonic interaction, even at low temperatures.
By linearly extrapolating to $T$=0
the measured dependence of the TO frequency on temperature (from Ref. \onlinecite{Bauer1978}) we obtain an unrenormalized
(harmonic) TO frequency of $\sim$13 cm$^{-1}$, rather close to that calculated with s-o interaction ($\sim$ 14 cm$^{-1}$, cf. Fig.\ref{fig3}).
This low TO frequency for $T\sim$ 0K underscores the nearly ferroelectric nature of this material
which would nominally become ferroelectric at $\sim$-70K.\cite{Bauer1978}

A disturbing feature of Fig \ref{fig3} is the dip in the LO-phonon band around the $\Gamma$-
point when s-o coupling is taken into account. We have been able to show
that the appearance of this dip is related to the closing and sign reversal of the electronic gap
induced by the s-o splitting of the valence and conduction band edges. Lowering the s-o
Hamiltonian by multiplication by an adjustable parameter
$\lambda \leq$ 1 (cf. Ref. \onlinecite{Diaz2007PRL}, see also Sect. IV) the electronic gap of
PbTe decreases from 0.32 eV for $\lambda$ = 1, becomes zero at  $\lambda$ $\sim$ 0.75 rising again  to 0.53 eV for $\lambda$ = 0
(the value of
the gap measured at 4 K is 0.19 eV (Ref. \onlinecite{Preier1979}). Note that for PbSe (Fig.2) a similar but smaller dip occurs when the s-o interaction is switched on.
In this case, the band gap nearly vanishes for $\lambda$ = 0 and becomes 0.52 eV for $\lambda$ = 1, with the same inverted order as in PbTe. Reversal of the gap is smaller than for PbTe because the s-o splitting of Se
is smaller than that of Te. For PbS (cf. Fig. \ref{fig1}) no dip is present other than that already existed without s-o splitting. The gap is $\sim$ 0 for $\lambda$ = 0 and 0.47 eV (also inverted) for $\lambda$ =1.

In view of the systematics of the effects just described we conjecture
that the anomalous dip in the LO frequencies observed at the $\Gamma$-point, is due to electron-phonon
interaction of the type  named after Kristoffel and Konsin.\cite{Kristoffel1972}  This interaction lowers the phonon
frequencies at $\Gamma$ and can even lead to ferroelectricity when a phonon frequency vanishes
(this is almost but not quite the case for PbTe but happens for SnTe, for which the gap is reversed with respect to PbTe.\cite{Tang1970,Cowley1997}
Since the measured LO frequency does not support the presence of the dip calculated for PbTe (and also less strongly for PbSe)
we conclude that the dip is due to some as yet unidentified effect of the s-o interaction,
probably related to the gap anomalies just described (and to the so-called gap problem ubiquitous
in LDA calculations). We have just made a guess as to how the LO band should look like once the problem
is removed, and represented it by a dotted red line, for both PbTe and PbSe (cf. Fig. \ref{fig2} and \ref{fig3}). Because of the small
volume of $k$-space encompassed by the dip ($\sim$ 0.1\% of the BZ) we believe that this problem
should not affect our calculations of heat capacities.

\section{Phonon Densities of States and Raman Spectra}

\begin{figure}[tbph]
\includegraphics[width=14cm ]{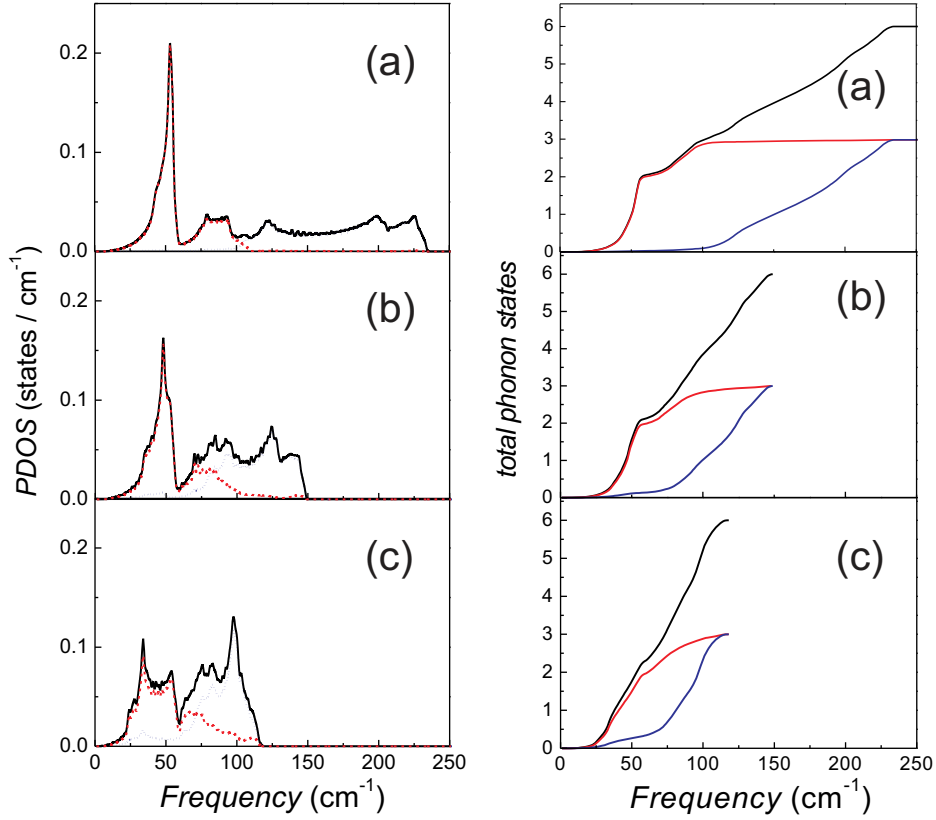}
\caption{(color online) (left) (a-c) Phonon density of states PDOS of PbS, PbSe and PbTe, from top to bottom, respectively.
The projections of the PDOS on the two constituents is shown by the dashed lines.
red: Pb projection; blue: projection on S, Se and Te, respectively.
(right) (a-c) Integrated PDOS (solid black line) showing the total number of states and their projections
on the cation (red line)  and the anion (blue line) constituents.}\label{fig4}
\end{figure}

The densities (DOS) of one-phonon states for the three materials under consideration
are displayed in Fig. \ref{fig4}, normalized to 6 states, as corresponds to one primitive cell (PC).
This figure shows not only the total DOS but also its projection onto each of the
two constituent atoms. A DOS calculated using the dispersion relations without s-o interaction
has already been reported for PbS in Ref. \onlinecite{Cardona2007}. Here we report only the calculations
performed with s-o interaction.

A comparison of the results in Fig. \ref{fig4} for PbS with those in Fig. \ref{fig2} of Ref. \onlinecite{Cardona2007}
(no s-o interaction)
clearly shows the shrinking of the width of the phonon bands, from 250 cm$^{-1}$  without s-o to $\sim$ 235 cm$^{-1}$ produced by the
s-o interaction. Also, whereas without s-o interaction the vibrations of the two constituents are almost fully separated
in frequency (dividing line 112 cm$^{-1}$), this is
not the case any more if s-o coupling is included. As shown in Fig. \ref{fig4} for PbS both
types of vibrations (Pb-like and S-like)
overlap in the region between 100 and 130 cm$^{-1}$, presumably related to the
shrinking of the phonon band width. In the case of PbSe three bands are seen in Fig. \ref{fig4}.
The low frequency one,  with a strong peak at $\sim$ 50 cm$^{-1}$, corresponds to Pb-like
vibrations of TA phonons, whereas the high frequency one, centered at
about 125 cm$^{-1}$, is LO- and TO- like and encompasses almost exclusively (95\%)
Se vibrations. The third band, centered around 80 cm$^{-1}$, is a mixture of LA- and TO- like modes
and encompasses vibrations of both constituent atoms with nearly the same amplitude.
The DOS of PbTe also exhibits 3 bands with a "pseudogap" at 60 cm$^{-1}$ and a second, not as
deep, at $\sim$ 85 cm$^{-1}$ .The lower band, mainly TA-like with some TO and LA contribution, is
predominantly Pb-like, whereas the middle band is about 66\% Te-like and 33\% Pb-like.
These facts, which will be useful when we discuss the dependence of the heat capacity on isotopic
mass, are quantitatively
represented in Fig. \ref{fig4} (right vignette) which displays the integrated total number of states and their projections,
as a function of frequency. The integrated (total) phonon states of PbS show clearly the almost
complete separation of Pb and S vibration, below 115 cm$^{-1}$ Pb-like, above S-like, and the existence
of a central band of  mixed modes in PbSe and PbTe.

\begin{figure}[tbph]
\includegraphics[width=8cm ]{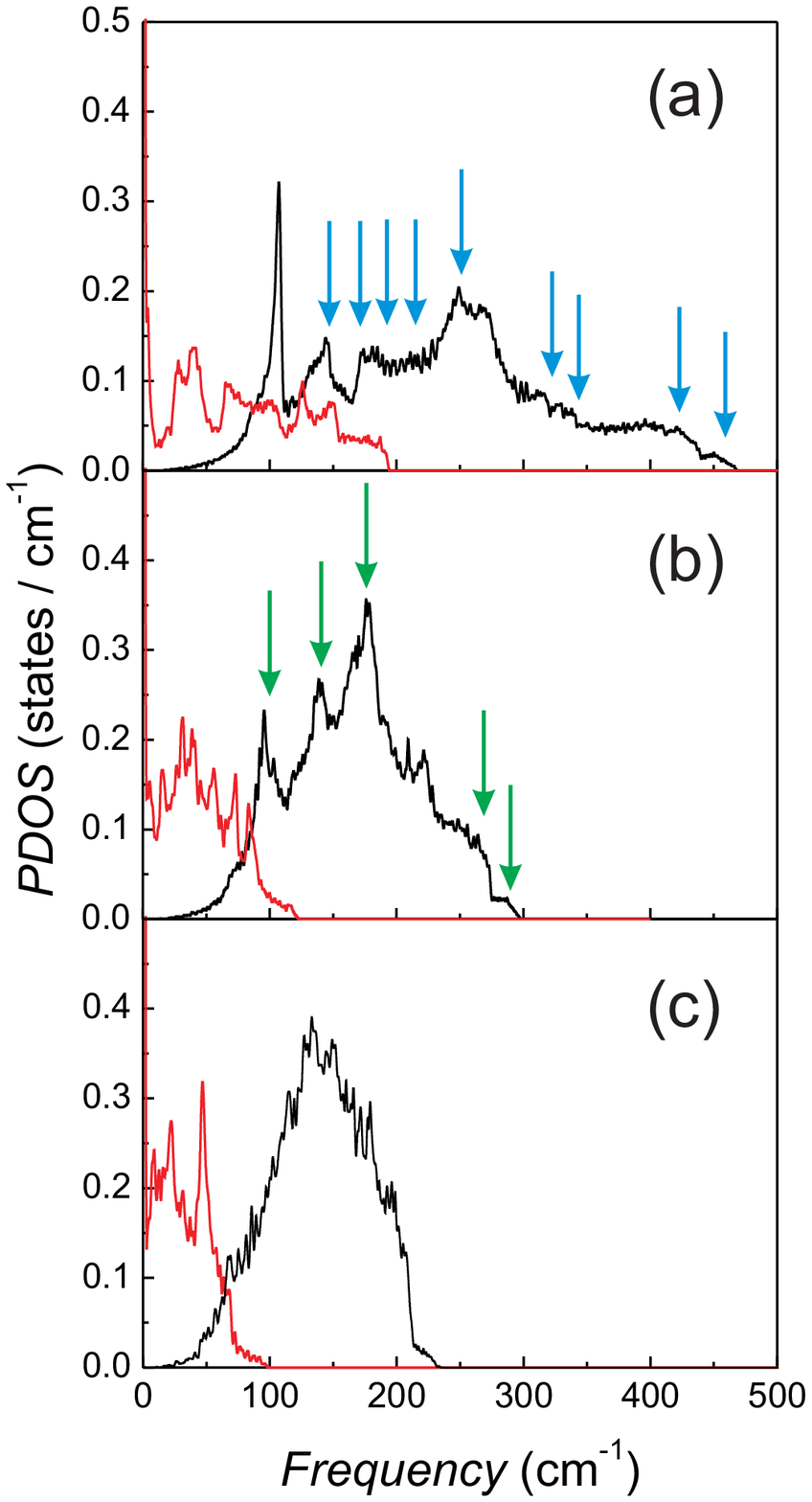}
\caption{(color online) (a-c) Densities of two-phonon states
(sum and difference) for PbS, PbSe, and PbTe, from top to bottom respectively,
with zero total $k$-vector.
Notice that for the sums the curves are normalized to 36 states (6$\times$6 states per
PC). For the differences they are normalized to 15 states (6$\times$5/2 staes per PC). The blue arrows mark
structure which is weakly (except for strong 430 - 460 cm$^{-1}$peaks) observed in the
Raman spectrum of PbS [ at  $\sim$ 150, 180, 205, 220, 250, 320, 340, 430, 460 cm$^{-1}$]. \cite{Etchegoin2008}
The green arrows indicate possible structure related to
the spectra published in Refs. \onlinecite{Chang2000} and \onlinecite{Ovsyannikov2004} for PbSe.}\label{fig5}
\end{figure}

Beside the one-phonon DOS just described, we have also calculated the DOS of two-phonon
states with total zero wave vector (or, equivalently, equal wave
vectors). These DOS are useful for the interpretation of second order Raman scattering
since first order Raman scattering is forbidden, cf. Ref. \onlinecite{Etchegoin2008}.
Two kinds of second order processes
are possible, one in which the frequencies of the two phonons
add (sum processes) and the other in which they subtract (difference processes). We calculate the DOS for both.
In Raman scattering sum processes are present even at $T$=0 and increase with
increasing temperature according to the appropriate statistical factors.
Difference processes have vanishing intensity at $T$=0 and can only be
observed with increasing temperature. We display in Fig. \ref{fig5} the densities of
two- phonon states for such sum and difference processes. Note that they are normalized
to 36 states (6$\times$6 states per PC) for the former and to 15 states (6$\times$5/2) for the latter, the factor of 2 corresponding to the separation of Stokes and anti-Stokes spectra.
The blue arrows mark structure which is weakly observed (except for the strong 430 - 460 cm$^{-1}$ peaks)
in the Raman spectrum of
PbS [ at $\sim$ 150, 180, 200, 220, 250, 320, 340, 430, 460 cm$^{-1}$].\cite{Etchegoin2008}
The green arrows indicate possible structure related to the spectra published
in Refs. \onlinecite{Chang2000} and \onlinecite{Ovsyannikov2004} for PbSe. No structure has been
identified as corresponding to phonon differences, although the band seen in Fig. \ref{fig5} at $\sim$ 40 cm$^{-1}$ ,
which seems to correspond to differences of TO and LA phonons (see Fig. \ref{fig1}), should increase strongly
with temperature and be identifiable with a good spectrometer that covers, with little straight light,
the corresponding region.	

Second order Raman spectra of PbSe and PbTe are rather scarce in the literature.
Some data are available for PbSe films grown on BaF$_2$ (Ref. \onlinecite{Chang2000})
and for bulk PbSe (Ref. \onlinecite{Ovsyannikov2004}) with a small
amount of Sn replacing Pb.  A doublet is observed at 265 - 300 cm$^{-1}$  which corresponds to
structure in the sum spectrum of Fig. \ref{fig5} (see (green) arrows) an is rather similar to the
doublet described above for PbS (430 - 460 cm$^{-1}$). The (green) arrows at 100, 135, 175, 265 and 290 cm$^{-1}$
also correspond to structure seen in the experimental spectra, although it is not clear whether
the 135cm$^{-1}$ structure is due to scattering by two phonons or forbidden scattering by one
LO phonon.\cite{Etchegoin2008}
A peak is also observed in the measured spectra at $\sim$ 90 cm$^{-1}$ (Ref. \onlinecite{Chang2000}).
It has been attributed
by the authors of Ref. \onlinecite{Chang2000} to difference scattering (LO-TO) although,
according to Fig.\ref{fig5}, it could
also be related to the peak in the sum DOS. Without measurements of the temperature
dependence of this peak it is not possible to clarify the assignment.  The Raman data
available for PbTe seem to be of poor quality because of segregation at
the surface of TeO$_2$  and other compounds:  The only structure that is clearly
identifiable corresponds to a single LO phonon ($\sim$120 cm$^{-1}$).\cite{Etchegoin2008b}

\section{Heat Capacities}

\begin{figure}[tbph]
\includegraphics[width=8cm ]{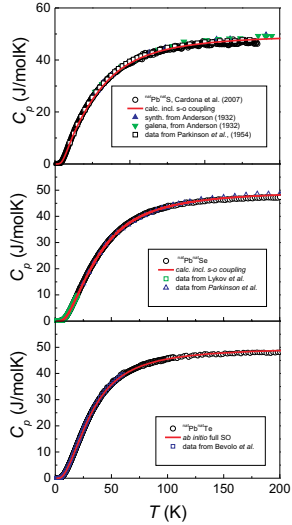}
\caption{(color online) Temperature dependence of the heat capacities of PbX (X=S, Se, Te) from top to bottom, respectively
measured
by us and other authors.
The results of our \textit{ab initio} calculations, which include s-o interaction, are shown by (blue) solid lines.
Literature data have been taken from Refs. \onlinecite{Anderson1932,Parkinson1954,Lykov1982,Bevolo1976}, as indicated.}\label{fig6}
\end{figure}

Figure \ref{fig6} displays the heat capacities of PbX (X = S, Se, Te) with
the natural isotope composition on either constituent.
For comparison, also literature data are shown.
We have chosen to plot the measured $C_p$ (and the calculated $C_v$ ) so as to be
able to read the data from the plots above $\sim$ 50 K with some accuracy. Below $\sim$ 50 K the
heat capacities become rather small and it is more convenient to plot $C_{p,v}/T^3$,
as will be done in the subsequent figure. Notice that at the highest temperatures in Fig. \ref{fig6},
the heat capacities tend to the Petit-Dulong limit ( $\sim$50 J/mol K).
The agreement between experimental and calculated results is excellent
although it is not possible, in these plots, to see the advantage of using in the
calculations a Hamiltonian with s-o interaction. Such advantage only appears below 50 K, especially
in the region where $C_p/T^3$ has a maximum.

\begin{figure}[tbph]
\includegraphics[width=10cm ]{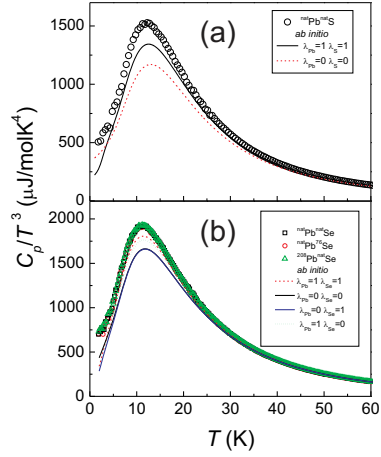}
\caption{(color online) (a) Heat capacity of PbS divided by $T^3$. The (black) circles represent our experimental
data (from Ref. \onlinecite{Cardona2007}). The (black) solid lines represents the \textit{ab initio} calculations
with s-o interaction for Pb and S, respectively ($\lambda_{Pb}$=$\lambda_{S}$=1). The red (dashed) line represents
the \textit{ab initio} calculation with s-o interaction  left out ($\lambda_{Pb}$=$\lambda_{S}$=0).
Note that for the calculations without s-o the maximum lies 27 \% below the measured one. Inclusion of
s-o interaction reduces the discrepancy by a factor of two. For the three curves (experimental
and \textit{ab initio})  the maxima occur at $\sim$12.5 K.
(b) Measured $C_p/T^3$ for three PbSe samples with different isotopic compositions as explained in the inset.
The heat capacities of the three samples with different isotope composition are almost indistinguishable.
The results of the \textit{ab initio} calculations are represented by the (red) dashed, the (green) dotted
and the (blue) and (black) solid lines, the latter two are almost indistinguishable, i. e. the effect of s-o
interaction of only Se is not noticeable, while there is a significant increase near the maximum
of $C_p/T^3$ ($T_{max}$ $\sim$ 11.3 K) if s-o interaction for Pb is taken into account ((red)
dashed and green (dotted) lines. For more details see inset).}\label{fig7}
\end{figure}

Figure \ref{fig7} displays the temperature dependence of $C_{v,p}/T^3$
as measured in the  2 - 60 K range and calculated with and without s-o interaction for PbS and PbSe, all
constituents with the natural isotope abundance. The data for PbS are identical to those published in Ref. \onlinecite{Cardona2007}.

It is of interest to compare the effect of the s-o interaction on the calculations of Fig. \ref{fig7}
with that found for bismuth (Ref. \onlinecite{Diaz2007PRL}) which
has a s-o interaction similar to that of lead.
The maximum in $C_v/T^3$ calculated for Bi without s-o interation is 20\% below the measured one.
Inclusion of s-o interaction yields a maximum 7\% higher than the measured one, i.e., 27\% higher
than that calculated without s-o interaction.\cite{Diaz2007PRL} The strong s-o effect calculated for
Bi may result from the fact that there are two equal atoms per PC of Bi whereas for the lead chacogenides there is only one
heavy atom (Pb) per PC.

Figure \ref{fig7}(b) displays $C_p/T^3$ as measured for three samples of PbSe with different
isotopic compositions, together with  three curves for $C_v/T^3$ calculated with
different contributions of the s-o interaction and with no s-o interaction.
The purpose of this exercise was to identify  the separate effects of the
s-o interactions of Pb (atomic s-o splitting of the 6$p$ electrons $\Delta_{\rm Pb}$= 1.27 eV)
and Se ($\Delta_{\rm Se}$ =0.42 eV (Ref. \onlinecite{Herman1963})).  The calculations in Fig. \ref{fig7}(b) correspond to full s-o interaction,
interaction only for Pb and interaction only for Se. Notice that the calculation with
only the s-o coupling of Se is rather close to that with no s-o coupling at all.
The s-o effect on $C_v$  is roughly proportional to the square of the s-o coupling parameter (cf. Ref. \onlinecite{Diaz2007PRL})
which for the valence electrons of Pb is three times bigger than that of Se. Hence, the effect
of the s-o of Se alone is expected to be $\sim$ 10 times smaller than that of Pb, i.e. unnoticeable
in Figure \ref{fig7}. However, $C_v/T^3$  with  s-o coupling for both constituents is slightly larger ($\sim$3\%)
than if only the s-o interaction of Pb is included.
This suggests the presence of bilinear terms in the corresponding perturbation
expression. In order to quantify these effects we define three partial contributions
$c_{\rm Pb}$ , $c_{\rm Se}$ and  $c_{\rm Pb-Se}$,  which represent the contribution to the s-o effect
quadratic in the s-o of Pb, in that of Te, and the bilinear contribution.
A fit of these effects on   $C_v/T^3$ shown in Fig. \ref{fig7}(b) yields:

\begin{equation}
c_{\rm Pb} = 99.6 \mu \rm{J/molK^4},     c_{\rm Se} \sim -3 \mu \rm{J/molK^4},
c_{\rm Pb-Se} = 76.2 \mu \rm{J/molK^4}\label{eq3}
\end{equation}

\begin{figure}[tbph]
\includegraphics[width=10cm ]{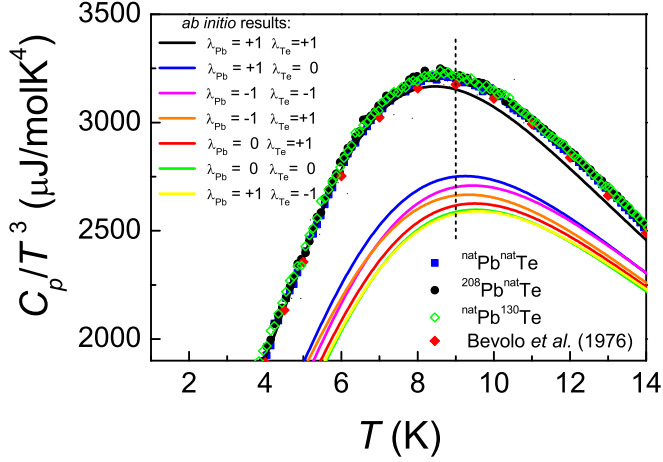}
\caption{(color online) (a) Heat capacity of PbTe divided by $T^3$. The symbols represent our
experimental data for various isotope compositions (cf. lower inset). The (black) solid line represents the \textit{ab initio}
calculations with s-o interaction for Pb and Te, respectively ($\lambda_{\rm Pb}$=$\lambda_{\rm Te}$=1). The (colored) solid lines
represent the \textit{ab initio} calculations with various combinations of the s-o interaction included (cf. upper inset, from top to bottom in the same order as in the figure). Literature data have been taken from Ref. \onlinecite{Bevolo1976}), as indicated.}\label{fig8}
\end{figure}

Obviously,  the bilinear term $c_{\rm Pb-Se}$ plays a rather important role
in the perturbation expansion, almost as important as the quadratic term.
This suggests the detailed investigation of a material with a larger s-o
coupling for the anion, such as PbTe  ($\Delta_{\rm Te}$ =0.86 eV, Ref. \onlinecite{Herman1963}).
The temperature dependence
of $C_p/T^3$ measured for samples with several isotopic compositions in the 3 - 14 K
region, where the maximum occurs, is shown in Fig. \ref{fig8}. As already suspected, bilinear
effects are large. They can be elucidated by multiplying the s-o Hamiltonian corresponding
to the two constituents by two parameters $\lambda_{\rm Pb}$ and $\lambda_{\rm Te}$, respectively.  Full s-o interaction is
obtained for $\lambda_{\rm Pb}$ = $\lambda_{\rm Te}$= +1.
Information about bilinear (and also cubic, cf. Ref. \onlinecite{Diaz2007PRL}) terms is readily
obtained by reversing the sign of one of either  $\lambda_{\rm Pb}$ or $\lambda_{\rm Te}$ . Additional information
concerning the perturbation expansion of $C_p/T^3$ versus s-o interaction is obtained
by setting one of either $\lambda_{\rm Pb}$ or $\lambda_{\rm Te}$  equal to zero, the other equal to one.

Figure \ref{fig8} shows  the experimental points for $C_p/T^3$ of PbTe obtained
with the isotopic compositions given in the inset. The vertical scale
is not wide enough to see differences for the various isotopes: such
differences will be discussed later. The experimental maximum occurs
at $\sim$ 8.5 K, nearly independently of isotopic composition. In order to
check the separate effects of $\Delta_{\rm Pb}$  and $\Delta_{\rm Te}$  on the \textit{ab initio}
calculations of $C_v/T^3$ we have multiplied the s-o interaction of
Pb and Te by $\lambda_{\rm Pb}$ and $\lambda_{\rm Te}$ , respectively.
We have then performed calculations for the six (after subtracting $\lambda_{\rm Pb}$ = $\lambda_{\rm Te}$ = 0)
sets of values of these separate parameters, as was done in the case of bismuth
in Ref.\onlinecite{Diaz2007PRL} (only for a single $\lambda$ corresponding to bismuth,
since there we dealt with a monatomic
crystal). The seven sets of  values of $\lambda_{\rm Pb}$ and $\lambda_{\rm Te}$ used, also given in the inset
of Fig. \ref{fig8}, are: (1,1), (1,-1), (-1,1), (-1,0) , (1,0), (-1,-1) and (0,0) for
($\lambda_{\rm Pb}$,$\lambda_{\rm Te}$). We have fitted
the values of $C_v/T^3$ at 9K (the vertical line in Fig. \ref{fig8}), after subtraction of the
value for $\lambda_{\rm Pb}$ = $\lambda_{\rm Te}$) = 0 (i.e., without s-o interaction for both constituents)
to four parameters chosen as follows:

\begin{equation}
C_v/T^3 (\lambda_{\rm Pb},\lambda_{\rm Te}) - C_v / T^3
(\lambda_{\rm Pb}=0,\lambda_{\rm Te}=0)= \lambda^2_{\rm Pb}c_1
+ \lambda^2_{\rm Te}c_2 +   \lambda_{\rm Pb}\lambda_{\rm Te}c_3   +   \lambda^3_{\rm Pb}c_4
\label{eq4}
\end{equation}

With five calculated values of Eq.(\ref{eq4}) and four fitting parameters
the corresponding values of $c_i$ are overdetermined and only an approximate fit is expected.

By minimizing the
corresponding variance we find the values of $c_i$ given in Table \ref{table1}
together with the calculated values of
$C_v/T^3$ ($\lambda_{\rm Pb}$ ,$\lambda_{\rm Te}$) and the fitted ones.
The fit with four parameters
of the five data seems to be  reasonable and thus the perturbation
expansion of Eq.(\ref{eq4}) is acceptable. Like in the case of PbSe discussed
above, the quadratic term $c_1$ is close to the bilinear one $c_3$ .
Also like in the case of Bi, with a s-o 6$p$  splitting similar to that
of Pb (Ref. \onlinecite{Herman1963}), a cubic term is needed to represent the calculated results.\cite{Diaz2007PRL}

\begin{table*} 
\caption{Expansion, $c_1 \lambda_{Pb}^2$ + $c_2 \lambda_{Te}^2$ +
$c_3 \lambda_{Pb}\lambda_{Te}$ + $c_4 \lambda_{Pb}^3$ used to quantify
the contributions to the quantity $C_p/T^3$(9K) if s-o coupling for either Pb ($\lambda_{Pb}$=1)
or Te ($\lambda_{Te}$=1) is taken or not taken ($\lambda_{Pb}$=$\lambda_{Te}$=0) into account (cf. Eq. (\ref{eq4})).
The values of the coefficients $c_i$ that optimally fit the data are:
$c_1$=247.5$\mu$ J/molK$^4$, $c_2$=64.2$\mu$ J/molK$^4$,
$c_3$=287.25$\mu$ J/molK$^4$, and $c_4$=-50.62$\mu$ J/molK$^4$.\label{table1}}
\begin{tabular}{ c |  c  c }
\hline\hline
$c_1 \lambda_{Pb}^2$ + $c_2 \lambda_{Te}^2$ + & \multicolumn{2}{c} {$\Delta C_p/T^3$(9K)}($\mu$ J/molK$^4$)\\
 $c_3 \lambda_{Pb}\lambda_{Te}$ + $c_4 \lambda_{Pb}^3$& calcul.  & experim.       \\
\hline
$c_1$ + $c_2$ + $c_3$ + $c_4$ & 548.3 & 565.7\\
$c_1$  + $c_4$ & 196.9 &   162.0\\
$c_2$ & 64.2 & 29.3\\
$c_1$ + $c_2$ - $c_3$ + $c_4$ & -26.2 & -8.8\\
$c_1$ + $c_2$ - $c_3$ - $c_4$ & 75.0 & 75.0\\
\hline
\end{tabular}
\end{table*}

\section{Dependence of the Heat capacities on the Isotopic Mass}

\begin{figure}[tbph]
\includegraphics[width=10cm ]{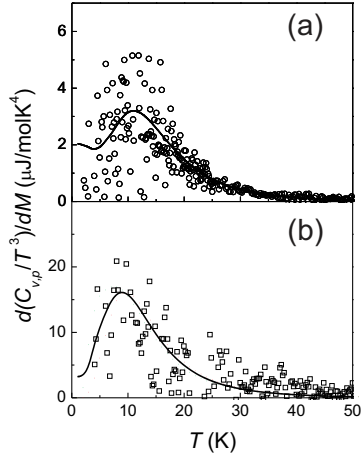}
\caption{Derivatives of $C_{v,p}/T^3$ of PbSe with respect to the isotopic
masses of (a) Pb and (b) Se  as measured using samples of different isotopic
compositions, compared with \textit{ab initio} calculations for various isotope
masses with s-o interaction included.}\label{fig9}
\end{figure}

The derivatives of $C_{v,p}/T^3$ with respect to the isotopic masses
of either Pb or S in PbS have been presented in Ref. \onlinecite{Cardona2007}, obtained
experimentally and by \textit{ab initio} calculations. Although the
calculations were performed without s-o coupling, the scatter of the
experimental data, due to the small range of isotopic masses available,
does not warrant a repetition of the calculations with s-o coupling.
We shall present here similar results for PbSe and PbTe, with the
\textit{ab initio} calculations including s-o interaction. Figure \ref{fig9} displays
the results obtained for PbSe using the isotopic samples given in the inset.

\begin{figure}[tbph]
\includegraphics[width=10cm ]{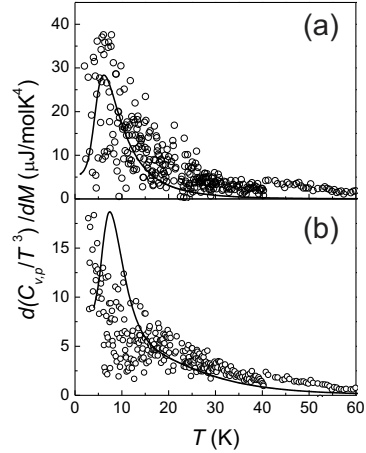}
\caption{Derivatives of $C_v/T^3$ of PbTe with respect to the isotopic
masses of (a) Pb and (b) Te  as measured using samples of different isotopic compositions,
compared with \textit{ab initio} calculations for various isotope masses
with s-o interaction included.}\label{fig10}
\end{figure}

A model for relating the position in the maxima of Figs. \ref{fig9} and \ref{fig10} was presented
in Ref. \onlinecite{Cardona2007}. It made use of a single Einstein oscillator to represent prominent
peaks in the one-phonon DOS (Fig. \ref{fig4}). The peak of the
mass derivatives (Figs. \ref{fig9} and \ref{fig10}) takes place at the Einstein
frequency (converted into temperature by multiplying by 1.44) divided
by 6.25. Using this procedure, we find for the maximum in Fig. \ref{fig9}  at 9 K ($M_{Pb}$ derivative)
an Einstein frequency of 40 cm$^{-1}$ which corresponds fairly well to the strongest peak seen
in Fig. \ref{fig4} for PbSe. For the Se derivative we find an Einstein frequency of 90 cm$^{-1}$,
in good agreement with the Se-like peak seen in Fig. \ref{fig4}.
Using the same procedure for Fig. \ref{fig10} we find from the $M_{Pb}$
derivative the Einstein frequency of 28 cm$^{-1}$, which also corresponds
rather well to the sharp Pb-like peak of Fig. 4 for PbTe. For the $M_{Te}$ derivative
we estimate an Einstein frequency of 38 cm$^{-1}$, which agrees with the
maximum in the Te projection of the DOS in the  corresponding vignette
of Fig 4 (lowest band of PbTe). This contribution of Te to the mass
derivative in the accoustic branch occurs because of the considerable
contribution of the anion vibrations in this region, as displayed
in Fig. \ref{fig4}. A similar anion contribution is negligible in the cases of PbSe and PbS.

\begin{figure}[tbph]
\includegraphics[width=10cm ]{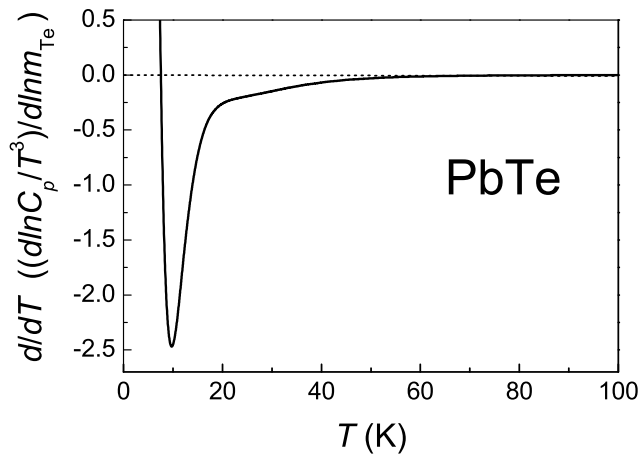}
\caption{(color online) Derivative with respect to $T$ of the calculated curve of Fig. \ref{fig10} (lower vignette),
showing a weak band in the 20-40 K region which corresponds to the Te-like band in Fig. \ref{fig4} for PbTe. See text.}\label{fig11}
\end{figure}

Since most of the Te weight to the DOS is in the optical phonon
region (90 cm$^{-1}$, Fig. \ref{fig4} ) there should also be a corresponding peak or band in the lower
vignette of Fig. \ref{fig10}. The experimental data in this region scatter
too much to allow observation of this peak. The calculated curve, however,
should exhibit some structure in this region. Although it is hard to
see in Fig. \ref{fig10} it is possible to detect it if the derivative with
respect to $T$ of the mass derivative is plotted. This is done in Fig. \ref{fig11}.
In the lower vignette of this Figure, concerning the $M_{Te}$ , we see indeed
weak structure between  20 and 40 K which corresponds to the
upper PbTe band of Fig. \ref{fig4} (90cm$^{-1}$$\times$1.44/6.2= 21K).

\section{Temperature dependence and isotopic mass dependence of \textbf{$C_{v,p}/T^3$}}

\begin{figure}[tbph]
\includegraphics[width=10cm ]{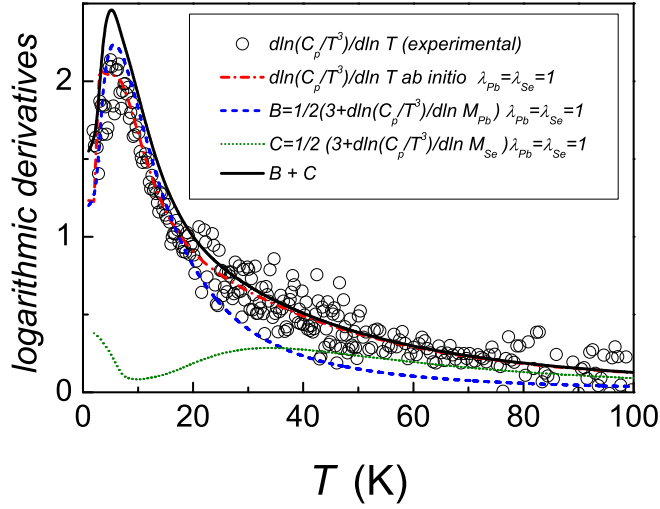}
\caption{(color online) Illustration of the relationship between the temperature dependence of $C_{v,p}/T^3$
and its dependence on the isotopic masses of both constituent atoms, drawn from \textit{ab initio} calculated data for
PbSe with natural isotope abundance for Pb and Se, respectively. The (black) circles represent experimental data for
the temperature dependence
($dln (C_{v,p}/T^3$)/$dlnT$), the (black) solid line represent the l.h.s of Eq. (\ref{eq2}). The other quantities are
explained in the inset.}\label{fig12}
\end{figure}

As already mentioned, the dependence of $C_v/T^3$ and $C_p/T^3$ on isotopic mass, or more precisely
the corresponding derivatives, can be obtained from the temperature dependence
of $C_{v,p}/T^3$ in the case of monatomic materials. The corresponding relation for diatomic
materials has been given in Eq. (\ref{eq2}). The correctness of this equation has been
demonstrated in Fig. 10 of Ref. \onlinecite{Cardona2007} for PbS. In Fig. \ref{fig12} we present one further example corresponding to PbSe.

The (red) dashed-dotted curve represents the  derivative of the calculated logarithmic
derivative of $C_v/T^3$ versus $T$. The green and blue curves represent the corresponding isotopic
mass derivatives as found in Eq. (\ref{eq2}) (obtained from the calculated data ($\lambda_{\rm Pb}$ = $\lambda_{\rm Te}$ = 1). The experimental data scatter
too much to obtain meaningful results). The black curve is the sum of these derivatives.
According to Eq. (\ref{eq2}), the black and red curves should be indistinguishable. This is indeed the
case  for $T >$ 30K. At lower temperatures the scatter of calculated and measured
points becomes rather large because both $C_{v,p}$   and $T^3$ are very small. Nevertheless
reasonable agreement is found. It is interesting to note that a change in the Pb mass
strongly affects the $T$ dependence of $C_v/T^3$   around 5 K whereas the change in the Se
mass affects that $T$ dependence in the 30-50 K region.

\section{Conclusions}
We have presented theoretical and experimental investigations of some lattice properties of three lead chalcogenides (PbS, PbSe, PbTe) with rock salt struture. Because of the presence of lead, these materials provide an excellent
case study for the elucidation of the effects of spin-orbit interaction on lattice properties as derived from \textit{ab initio} electronic structure calculations. Among other properties that have been investigated are the lattice parameters, the phonon dispersion relations, the heat capacity and their dependence
on isotopic masses. The effect of spin-orbit interaction on the lattice parameters is less than 1\% and can be neglected compared with other sources of computational error. Its effect on the lattice dynamics and the corresponding low temperature heat capacities is considerable. For these properties spin-orbit interaction
significantly improves the agreement of theory with experiment.
Because of the large scatter in the effect of isotope masses on the heat capacity, related to the small range of
of isotope mass available, spin-orbit interaction does not play an important role in the comparison of
theoretical and experimental data for the isotope mass dependence of the heat capacity. This dependence can be modeled on the basis of two Einstein oscillators.

We have calculated, from the \textit{ab initio} phonon dispersion relations, the corresponding density of one-phonon states and those of two-phonon states relevant for optical spectroscopies. The latter have been compared with few extant second order Raman spectra of these materials.

\begin{acknowledgments}
We gratefully acknowledge the computer
time allocation at the Centro Nacional de Superc$\acute{\rm o}$mputo, IPICyT.
We thank K. Graf for assistance with data processing and M. Krack for help with the
generation of pseudopotentials. J.S. acknowledges financial support of a Spanish CICYT grant MAT2007-60087 and Generalitat de Catalunya grants 2005SGR00535 and 2005SGR00201. A. H. R. acknowledges support from CONACIT (Mexico) under project J-59853-F.

\end{acknowledgments}

\end{document}